\documentclass{article}

\usepackage{arxiv}

\usepackage[utf8]{inputenc} 
\usepackage[T1]{fontenc}    
\usepackage{hyperref}       
\usepackage{url}            
\usepackage{booktabs}       
\usepackage{amsfonts}       
\usepackage{nicefrac}       
\usepackage{microtype}      
\usepackage{lipsum}		
\usepackage{graphicx}
\usepackage{doi}

\title{On the Approximability of External-Influence-Driven Problems}

\author{ \href{https://orcid.org/0000-0000-0000-0000}{\includegraphics[scale=0.06]{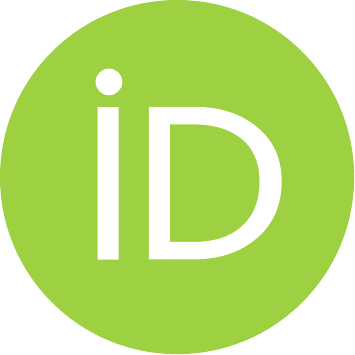}\hspace{1mm}Panagiotis Aivasiliotis} \\
	School of Electrical and Computer Engineering\\
	National Technical University of Athens\\
	Greece \\
	\texttt{p.aivasiliwths@gmail.com} \\
	\And
	\href{https://orcid.org/0000-0002-6220-3722}{\includegraphics[scale=0.06]{orcid.pdf}\hspace{1mm}Aris Pagourtzis} \\
	School of Electrical and Computer Engineering\\
	National Technical University of Athens\\
	Greece \\
	\texttt{pagour@cs.ntua.gr} \\
}

\date{}

\renewcommand{\headeright}{}
\renewcommand{\undertitle}{}
\renewcommand{\shorttitle}{On the Approximability of External-Influence-Driven Problems}

\hypersetup{
pdftitle={A template for the arxiv style},
pdfsubject={q-bio.NC, q-bio.QM},
pdfauthor={David S.~Hippocampus, Elias D.~Striatum},
pdfkeywords={First keyword, Second keyword, More},
}

\usepackage{mathtools}
\usepackage{amsthm}
\usepackage{amsfonts}
\usepackage{algorithm}
\usepackage{algorithmic}

\usepackage{amsmath}
\usepackage[nameinlink]{cleveref}

\newtheorem{definition}{Definition}
\newtheorem{theorem}{Theorem}
\newtheorem{lemma}{Lemma}
\newtheorem{corollary}{Corollary}[theorem]
\newtheorem{claim}{Claim}
\newtheorem{observation}{Observation}
\newtheorem{proposition}{Proposition}
\newtheorem{remark}{Remark}

\DeclareMathOperator*{\argmax}{arg\,max}

\DeclarePairedDelimiter\floor{\lfloor}{\rfloor}

\begin{document}
\maketitle

\begin{abstract}
Domination problems in general can capture situations in which some entities have an effect on other entities (and sometimes on themselves). The usual goal is to select a minimum number of entities that can influence a target group of entities or to influence a maximum number of target entities with a certain number of available influencers. In this work, we focus on the distinction between \textit{internal} and \textit{external} domination in the respective maximization problem. In particular, a dominator can dominate its entire neighborhood in a graph, internally dominating itself, while those of its neighbors which are not dominators themselves are externally dominated. We study the problem of maximizing the external domination that a given number of dominators can yield and we present an~$0.5307$-approximation algorithm for this problem. Moreover, our methods provide a framework for approximating a number of problems that can be cast in terms of external domination. In particular, we observe that an interesting interpretation of the maximum coverage problem can capture a new problem in elections, in which we want to maximize the number of \textit{externally represented} voters. We study this problem in two different settings, namely Non-Secrecy and Rational-Candidate, and provide approximability analysis for two alternative approaches; our analysis reveals, among other contributions, that an earlier resource allocation algorithm is, in fact, a 0.462-approximation algorithm for maximum external domination in directed graphs.
\end{abstract}


\section{Introduction}
\label{sec:intro}
The notion of influence has been fundamental in social networks, occupying a plethora of studies that focus on various influence optimization objectives. An arguably favorite setting of influence optimization is that of vertex domination problems. The usual goal is to minimize the number of influencers needed to influence the whole network, assuming certain patterns of influence (e.g. immediate neighborhood, or $k$-hop neighborhood). Practically though, only a limited number of influencers is available, implying that we can influence only a part of the network. This gives rise to an optimization problem that asks to maximize the number of influenced entities given the number of available influencers. The above described problems comprise respectively the minimization~\cite{haynes1998fundamentals,ghoshal2017topics,10.1007/978-3-319-04126-1_12} and the maximization version~\cite{miyano2011maximum,jia2020blocking} of ($k$-hop~\cite{kutten1995fast}) dominating set in the (un)directed graph representing the network (assuming that edges represent some social acquaintance implying influence).  

In the context of influence maximization, when cast as some domination problem, it makes sense to consider the number of externally dominated entities, as this is a natural measure of the success of an influence attempt. Interestingly, to the best of our knowledge, the maximization of such an objective has not been studied in the literature, apart from particular special cases~\cite{fotakis2020object}, despite the huge volume of research on influence maximization and domination problems. Two main questions thus arise: How does this new objective differ compared to the standard one in terms of exact solvability? Second, when an approximate solution is solicited, what ratio we can guarantee? The second question is really intriguing as it concerns underrated yet significant information: the maximum number of externally influenced entities and the portion of them that can be guaranteed by an efficient approximation algorithm. 

Even though it is known that the best approximation that can be achieved for the standard (maximization) problems is $(1-1/e)$ (unless $\textbf{P} = \textbf{NP}$)~\cite{miyano2011maximum,10.1145/285055.285059}, the approximability of the new variants is far from being considered settled. The reason for the lack of relevant literature might be that, at first glance, the problem of maximizing the total number of dominated entities (including the dominators) in (un)directed graphs seems to be essentially identical to that of maximizing the number of externally dominated ones. Surprisingly, this is not always the case, especially when it comes to approximability, but also with respect to exact solvability, as can be demonstrated by certain examples. Besides, potential reductions to well-studied problems such as facility location~\cite{cornuejols1977exceptional,cornuejols1977uncapacitated,DBLP:journals/dam/AgeevS99}, do not seem to be fruitful as we will explain later. 

The problem of maximizing the external influence is meaningful not only in undirected graphs, used to express symmetric relations in social networks (such as friendship) but also in directed graphs, which can express non-symmetric relations (such as followers). 
Our work is primarily focused on two problems that capture these two cases and which we introduce here under the names Max ($k$-hop) Ext-Domination\footnote{In fact, the case $k=1$ is very closely related to a problem introduced in~\cite{fotakis2020object}, which can be considered as a first approach to capture the idea of external influence maximization.} and Max Ext-Representation respectively. The name of the latter (which in fact is a generalization of maximum external domination in directed graphs, a problem that can be defined similarly to the undirected version) reflects the fact that this problem also captures an election optimization objective, namely to elect a committee in a way that maximizes the externally represented voters, that is, voters who are not members of the elected committee.

\subsection{Our contribution}
We begin by studying the (in)approximability of Max $k$-hop Domination which is the maximization version of $k$-hop Dominating Set~\cite{kutten1995fast,10.1007/978-3-319-04126-1_12} showing that it can be approximated within $(1-1/e)$ and that for any fixed $k$, the problem is hard to approximate within $(1-1/e + \varepsilon)$ for any $\varepsilon > 0$, generalizing the known results for $k = 1$~\cite{miyano2011maximum}. These results, as well as the (in)approximability results for a variant of Max Coverage (obtained directly from Max Coverage), will be used in our effort to derive approximation guarantees for several algorithms addressing the Max $k$-hop Ext-Domination and Max Ext-Representation problems. 

In particular, we will present a framework for approximating Max $k$-hop Ext-Domination achieving a $(6e-5)/(6e+5)$-approximation ratio for Max Ext-Domination (which is the special case $k = 1$). Our results improve on the approximation ratio achieved in~\cite{fotakis2020object} for a problem of object allocations, called OPT-EXT(0,1).

We also show that Max Ext-Representation can be approximated within $(e-1)/(e+1)$, under a specific setting called Non-Secrecy, by appropriately adapting the analysis of an algorithm given in~\cite{fotakis2020object}. We also study the approximability of the greedy heuristic, which is especially useful in the Rational-Candidate Setting, where it seems harder---generally speaking---to derive approximation guarantees. Under the assumption that candidates should approve of at least one candidate other than themselves, the greedy heuristic can achieve the approximation ratio $(e-1)/(e+1)$. Such an assumption captures cases of social media networks (supporting non-symmetric relations), where all profiles are both potential promoters or receivers of promoted products, thus some sort of well-connected sub-network is preferred, guaranteeing for example that all profiles in a given instance should follow at least one other profile.

Note that, for the sake of readability, most proofs are deferred to the Appendix.

\subsection{Related work}\label{relatedWork}
We distinguish three different contexts within which we can formulate the Max Ext-Domination problem. Some formulations can actually give concrete results while others do not seem, from our understanding, to add to our analysis.

The first and broader context is that of submodular maximization. A set function $f : V \rightarrow \mathbb{R}$ is submodular if $f(A \cup \{e\}) - f(A) \geq f(B \cup \{e\}) - f(B)$ 
holds for every $A \subseteq B \subseteq V$ and $e \in V\backslash B$. If the previous inequality holds with equality then $f$ is called modular. Submodular maximization has been studied extensively in its many variations and the results vary depending for example on whether the objective function is monotone ($f$ is monotone if $A \subseteq B$ implies $f(A) \leq f(B)$, for any $A, B \subseteq V$) or on the particular constraints determining the set of feasible solutions. For example, cardinality constraints reduce the set of feasible solutions to $\{A \in V : |A| \leq p\}$ for some given integer $p$. The objective of Max Domination which is a maximum coverage special case is submodular and monotone. By excluding the internally dominated vertices, the objective sustains the submodularity but forfeits the monotonicity. The problem of maximizing a non-monotone submodular function subject to cardinality constraints is quite challenging and enjoys several notable results. Currently, the best approximation known is 0.385 due to~\cite{buchbinder2019constrained}. We also individuate the recent work of~\cite{kuhnle2019interlaced} and the earlier work of~\cite{buchbinder2014submodular}. In particular, two different cardinality constraints are considered in~\cite{buchbinder2014submodular}, to either set an upper bound on the cardinality of the solution or to require a particular cardinality. In general, these two cardinality constraints may imply two different problems, however, in our case (Max Ext-Domination), they are equivalent in the sense that we can easily reduce the problem with upper-bounded cardinality to the one with exact cardinality, and vice versa, while preserving the same approximation guarantees. 
Continuing in the same context, our objective apparently belongs to the family of the objectives being studied by~\cite{sviridenko2017optimal,harshaw2019submodular} since it can be expressed as the difference between a submodular and a modular function. Nevertheless, their results do not add to our analysis.

Narrowing down the context in which we address Max Ext-Domination, we can show that it can also be formulated as an instance of the uncapacitated facility location problem (UFLP)~\cite{cornuejols1977exceptional,cornuejols1977uncapacitated} but as it seems we cannot cherish the same approximation results that we have for UFLP. In particular, it can be shown that the approximation ratio (of any greedy algorithm) for Max Ext-Domination cannot match the ratio $(1-(1-1/p)^p)$ achieved by UFLP (consider a path of five vertices where two of them should be designated as the dominators, which is a counter-example used for an argument of the same nature in~\cite{fotakis2020object}).

The third and most practical context is that of formulating an instance of Max Ext-Domination as an instance of OPT-EXT(0,1), a problem of object allocations on a graph~\cite{fotakis2020object}. This formulation ensures a $(e-1)/(e+1)$-approximation algorithm due to~\cite{fotakis2020object}. 

Regarding Max Ext-Representation, its objective can be shown to be a non-monotone submodular function. Furthermore, a potential formulation of the problem as an instance of UFLP, at first glance, seems to fail and the reason is that the costs are not fixed, something that UFLP requires by definition. However, subject to certain conditions, which ensure that all candidates approve of themselves the formulation indeed succeeds; this can be either achieved by appropriate modifications to the input or it can be assumed to hold in particular settings. Yet, employing similar arguments to those above for Max Ext-Domination, which is also a special case of Max Ext-Representation, the problem cannot share the same approximability results with UFLP.

\paragraph*{Related work on inapproximability}
The maximization version of the dominating set problem has been studied thoroughly by~\cite{miyano2011maximum}. We refer to this problem as Max Domination, partially adopting the name they use in that work. A simple greedy algorithm is able to achieve a $(1-1/e)$-approximation guarantee which is also the best possible (unless \textbf{P} = \textbf{NP}). The inapproximability bound is shown with the help of a gap-preserving reduction (see~\cite{Vazirani2003}) from the Maximum Coverage Problem. 

\paragraph*{Additional related work on submodular maximization} It is known that it is impossible (unless \textbf{P} = \textbf{NP}) to achieve a 0.478-approximation for the problem of submodular maximization subject to cardinality constraints~\cite{gharan2011submodular}. Given this result, FPT time approximation algorithms have recently drawn attention. Among these attempts, we individuate the recent work of~\cite{rubinstein_et_al:LIPIcs.ICALP.2022.106}. Note that their 0.539-approximation FPT time algorithm is---to the best of our knowledge---the only result from the literature beating (slightly) our $(6e-5)/(6e+5)$-approximation ratio ($(6e-5)/(6e+5) \approx 0.530$) for Max Ext-Domination, however, that algorithm is an FPT one, while ours is a polynomial time algorithm.


\section{Coverage and domination maximization under the standard objective}
Coverage and in particular domination problems are known for their capability of capturing a notion of influence; the domination of a vertex or the representation of a voter are two examples of such influence. In this section, we deal with the following problems: Maximum Coverage under exact cardinality constraints (MCP-ECC) and Max ($k$-hop) Domination.  We study the (in)approximability of the two problems, the objectives of which can be regarded as `standard' objectives; the results we obtain will be later used in studying the approximability of variants of these problems the objectives of which deviate from the `standard' objectives, in an effort to capture a notion of \textit{external influence} as mentioned in \Cref{sec:intro}.     

\subsection{MCP-ECC}
\begin{definition}[Maximum Coverage under exact cardinality constraints] 
Given a set $X$ of $n$ elements, a collection $\mathcal{S} = \{S_1, \ldots, S_m\}$ of $m$ subsets of $X$, and an integer $l$, determine
\[\argmax_{\mathcal{S'} \subseteq \mathcal{S} : |\mathcal{S'}| = l} \Bigg\vert \bigcup_{S_i \in \mathcal{S}'}S_i\Bigg\vert.\]
\end{definition}

MCP-ECC considers a more strict constraint regarding the cardinality of feasible solutions compared to that of Maximum Coverage (MCP) where feasible solutions typically consist of at most $l$ sets. Nevertheless, it can be shown that the two problems share the same (in)approximability results. More specifically, MCP is known to be \textbf{NP}-hard~\cite{10.1145/285055.285059} and a simple greedy algorithm that picks $l$ sets so that the $i$-th set contains the maximum number of uncovered elements is able to guarantee an approximation ratio of $(1-1/e)$~\cite{hochbaum1996approximating} which is also the best we can hope for (unless \textbf{P} = \textbf{NP})~\cite{10.1145/285055.285059}.

As claimed above, these results also apply in the case of MCP-ECC. To see this, observe that the objective function, which counts the 
elements covered by a subcollection $\mathcal{S}' \subseteq \mathcal{S}$, is non-decreasing, hence any feasible solution for MCP with less than $l$ sets can be transformed into a feasible solution for MCP-ECC consisting of exactly $l$ sets without its objective value being decreased.

\subsection{Max \ensuremath{k}-hop Domination and Max Domination}
\begin{definition}[Max $k$-hop Domination and Max Domination]
Given an undirected graph $G = (V, E)$ and integers $p, k$, \textup{Max $k$-hop Domination} asks for designating a set of $p$ vertices as the \textup{dominators} so that the total number of vertices dominated by the dominators is maximized; a vertex is regarded as \textup{dominated} if either has been designated as a dominator or it is in a distance of at most $k$ hops from a designated dominator. 
For $k = 1$, we refer to the problem as \textup{Max Domination}\footnote{The case $k=1$ has been studied in~\cite{miyano2011maximum} where it is shown, among other results, to be $\textbf{NP}$-hard, which also settles the hardness of Max $k$-hop Domination.}.
\end{definition}

Regarding the approximability of Max $k$-hop Domination, we can easily formulate the problem as a special case of MCP-ECC and thus guarantee an approximation ratio of $(1-1/e)$ using the greedy algorithm. The total number of dominated vertices (given a set of dominators) is then given by the following function $dom$: 
\[dom(A) = \Bigg\vert\bigcup_{v \in A}\mathcal{N}_k[v]\Bigg\vert, A \subseteq V,\]
where $\mathcal{N}_k[v] = \{u \in V : 0 \leq dist(v,u) \leq k\}$ is the closed $k$-neighborhood of $v$ consisting of all vertices in a distance of at most $k$ edges from $v$ (including $v$).

When it comes to studying the inapproximability of the problem, we distinguish between the cases of $k$ being fixed and of being part of the input. It is known~\cite{miyano2011maximum} that for fixed $k = 1$, the problem has an inapproximability bound of $(1-1/e)$ (unless \textbf{P} = \textbf{NP}), hence when $k$ is not fixed the problem has the same inapproximability bound. We show that the same bound holds for any fixed $k \geq 1$ too but the proof requires establishing appropriate gap-preserving reductions, a tool that has been used in~\cite{miyano2011maximum} to show the bound ($1-1/e$) for $k = 1$. Our generalized result is presented in the next theorem (the definition of the graphs used in establishing the new gap-preserving reductions and the proof, are deferred to the appendix; some of the notations we employ have been (partially) adopted from~\cite{miyano2011maximum}). 
\begin{theorem}\label{thm:inapprox}
There is no polynomial time approximation algorithm with any constant factor better than $(1-1/e)$ for \textup{Max $k$-hop Domination} unless \textup{\textbf{P} = \textbf{NP}}, for any fixed integer $k \geq 1$.
\end{theorem}

\section{Max \ensuremath{k}-hop Ext-Domination and Max Ext-Domination}
\subsection{Problem Definition}
To instantiate the notion of external domination in Max $k$-hop Domination, we first need to distinguish between internally and externally dominated vertices, which will eventually give rise to a new variant, called Max $k$-hop Ext-Domination, which we study in the present section.
\begin{definition}[Max $k$-hop Ext Domination and Max Ext-Domination] The externally dominated vertices are those dominated vertices that are not dominators themselves, considering dominators as \textit{internally} dominated. The function $ext$, which counts the vertices being externally dominated, can be then defined as follows:
\[ext(A) = dom(A) - |A|, A \subseteq V.\]
We will refer to the problem of maximizing $ext$ given the number of dominators as \textup{Max $k$-hop Ext-Domination}, while for $k=1$ we will simply refer to the problem as \textup{Max Ext-Domination}. 
\end{definition}

Note that an exact algorithm for Max ($k$-hop) Domination implies an exact algorithm for Max ($k$-hop) Ext-Domination and vice versa, that is, with respect to a particular instance, the two problems share the same set of optimal solutions. However, when it comes to approximate solutions the situation is different; this calls for a study of the approximability of Max ($k$-hop) Ext-Domination.

To this end, we present a framework for approximating the entire family of Max $k$-hop Ext-Domination problems, providing concrete results that the framework yields. In fact, Max Ext-Domination is the main problem we are interested in since it finds direct applications to recent problems concerning resource allocations, to be presented later on.

\subsection{A framework for approximating Max \ensuremath{k}-hop Ext-Domination}\label{ss32}
As already mentioned, we can approximate Max $k$-hop Domination within a factor of $(1-1/e)$ using the greedy algorithm, to which we will refer as \textsc{Greedy}\footnote{The input of \textsc{Greedy} consists of the graph $G$ and the two integers $p, k$. Parameters $p$ and $k$ are usually clear from the context, hence we omit to refer to them.}, which produces a sequence of sets $(A_0, A_1,\ldots, A_p)$ such that $A_0 = \emptyset \text{ and } A_{i+1} = A_i \cup \{\argmax_{v \in V\backslash A_i}dom(A_i \cup \{v\})\}, i < p$.
Formally, we have $dom(A_p) \geq (1-1/e)dom(A_p^*)$, which we rewrite and get
\[ext(A_p) \geq (1-1/e)ext(A^*_p) - p/e,\]
where $A^*_p$ is an optimal solution (recall that Max $k$-hop Domination and Max $k$-hop Ext-Domination are equivalent in terms of exact solvability).

By defining $\theta_p \coloneqq \frac{ext(A_p)}{p}$ $(>$ $0)$ the former inequality becomes
\[ext(A_p) \geq \frac{\theta_p(e-1)}{1+\theta_p e}ext(A_p^*).\]
We also let $\sigma_p \coloneqq \frac{ext(A_p)}{n-p}$, $n = |V|$.
Since $ext(A_p) \geq \sigma_p\cdot ext(A_p^*)$, we finally deduce that
\[ext(A_p) \geq \max\{\frac{\theta_p(e-1)}{1+\theta_p e}, \sigma_p\} ext(A_p^*).\]
\begin{proposition}\label{prop:theta}
$\theta_1 \geq \theta_2 \geq \ldots \geq \theta_{n-1}$ and $\sigma_1 \leq \sigma_2 \leq \ldots \leq \sigma_{n-1}$.
\end{proposition}
The next claim follows.
\begin{claim}
Let $\theta, \sigma$ be numbers such that for any instance of Max $k$-hop Ext-Domination, there is some integer $d$ such that ${\theta}_p \geq {\theta}, \text{ for $p \leq d $} \text{ and } {\sigma}_p \geq \sigma, \text{ for $p > d$}.$\footnote{An accurate notation would require writing $\theta(k), \sigma(k)$ to underline the implicit dependence on parameter $k$. However, it is always clear from the context, hence we can abuse notation and write $\theta, \sigma$ instead.} \Cref{prop:theta} then indicates that a lower bound of $\min\{\frac{{\theta}(e-1)}{1+{\theta} e}, {\sigma}\}$
on the approximation ratio achieved by the greedy heuristic can be guaranteed for Max $k$-hop Ext-Domination.
\end{claim}

In fact, appropriately defined auxiliary graphs can be used so as to provide lower bounds similar to the ones defined above which is most useful as we are not limited by the possible complicacy of the graph $G$, which obviously affects how well one can determine the values $\theta, \sigma$. In particular, let $\widehat{G} = (\widehat{V}, \widehat{E})$ be a subgraph of $G$, where $\widehat{V} = V$ and $\widehat{E} \subset E$. We let $\widehat{dom}$ denote the function that counts the total number of dominated vertices, given a set of dominators, in the auxiliary graph $\widehat{G}$ and $\widehat{ext}$ denote the function that counts the externally dominated vertices (in $\widehat{G}$). It easily follows that $dom(A) \geq \widehat{dom}(A), A \subseteq V$.
Likewise, we let $(\widehat{A}_1,\ldots, \widehat{A}_p)$ be the result of \textsc{Greedy} when it is called for $\widehat{G}$ and $\widehat{\theta}_p, \widehat{\sigma}_p$ be defined similarly to $\theta_p, \sigma_p$. We can show that
\[
\max\{ext(A_p), ext(\widehat{A}_p)\} \geq \max\{ext(A_p), \widehat{ext}(\widehat{A}_p)\} \geq \max\{\frac{\widehat{\theta}_p(e-1)}{1+\widehat{\theta}_p e}, \widehat{\sigma}_p\}ext(A_p^*).
\]

In the same fashion as before, the next claim follows.
\begin{claim}
Let $\widehat{\theta}, \widehat{\sigma}$ be numbers such that for any instance of Max $k$-hop Ext-Domination (more precisely, for the corresponding auxiliary instances), there is some integer $\widehat{d}$ such that $\widehat{\theta}_p \geq \widehat{\theta}, \text{ for $p \leq \widehat{d}$ and } \widehat{\sigma}_p \geq \widehat{\sigma}, \text{ for $p > \widehat{d}$}$. Then a lower bound of $\min\{\frac{\widehat{\theta}(e-1)}{1+\widehat{\theta} e}, \widehat{\sigma}\}$ on the approximation ratio described above can be guaranteed for (any instance of) Max $k$-hop Ext-Domination. 
\end{claim}
Our goal is to maximize this lower bound as detailed below where we give concrete examples of auxiliary graphs and the corresponding lower bounds achieved using them.

\subsection{Auxiliary graphs and lower bounds}
Assuming connected graph $G$ first, we present in \Cref{lem:lowerbounds} the lower bounds achieved by an auxiliary graph in two different scenarios; the auxiliary graphs are produced by \Cref{alg1} for any spanning tree $T$ of $G$ and for each one of the two possible values of $\delta \in \{0,1\}$. Using these results, in \Cref{lem:gen} we show the approximation ratio achieved by \Cref{alg2} for Max $k$-hop Ext Domination for any graph $G$.

\begin{algorithm}[htb]
\caption{\textsc{Decomposition}}\label{alg1}
\textbf{input}: A rooted tree $T$ and parameters $\delta \in \{0,1\}$, $k$.\\
\textbf{output}: The tree $T$ or the union of $t$ rooted minimal trees, each of at least $k+1+\delta$ vertices.

\begin{algorithmic}[1]
\IF{$|V(T)| < k + 1 + \delta$}
\RETURN{$T$}
\ENDIF
\STATE{$i \leftarrow 0$}
\WHILE{$T \neq \emptyset$}
\STATE{$i \leftarrow i + 1$}
\STATE{Choose a vertex $u_i$ such that the subtree that is formed by vertex $u_i$ and all its descendants has at least $k + 1 + \delta$ vertices and is minimal, which means that each of its subtrees has at most $k+\delta$ vertices.}
\STATE{Store the subtree formed by $u_i$ and its descendants to $\widehat{G}_i$.}
\STATE{Label vertex $u_i$ as the root of $\widehat{G}_i$.}
\IF{$0 < |V(T\backslash\widehat{G}_i)| < k + 1 + \delta$}
\STATE{$\widehat{G}_i \leftarrow T$ /* \textit{Vertex $u_i$ labeled as the root of $\widehat{G}_i$ in step 9 remains the root */}}
\ENDIF
\STATE{$T \leftarrow T \backslash \widehat{G}_i$}
\ENDWHILE
\RETURN{$\bigcup_{i=1}^{t}\widehat{G}_i$}
\end{algorithmic}
\end{algorithm}

\begin{algorithm}[htb]
\caption{}\label{alg2}
\textbf{input}: A graph $G$, the number $p$ of dominators, the number $k$ of hops and $\delta \in \{0,1\}$.

\textbf{output}: A set $D$ of $p$ vertices that have been designated the dominators.

\begin{algorithmic}[1]
\STATE{$F \leftarrow \text{\textsc{SpanningForest}}(G)$}
\STATE{$\widehat{G} \leftarrow \emptyset$}
\FOR{\textbf{each} tree $T \in F$}
\STATE{$\widehat{G} \leftarrow 
\widehat{G}$ $\bigcup$ \textsc{Decomposition}($T, \delta, k$)}
\ENDFOR
\STATE{$A_p \leftarrow$ \textsc{Greedy}($G, p, k$)}
\STATE{$\widehat{A}_p \leftarrow$ \textsc{Greedy}($\widehat{G}, p, k$)}
\RETURN{$\argmax_{D \in \{A_p, \widehat{A}_p\}}ext(D)$}
\end{algorithmic}
\end{algorithm}

\begin{lemma}\label{lem:lowerbounds}
Using the auxiliary graph produced by \textsc{Decomposition}$(T, 0, k)$ we get $\widehat{\theta}_p \geq k, p \leq t$ and $\widehat{\sigma}_p = 1, p \geq t$ yielding a lower bound of $\frac{e-1}{e+\frac{1}{k}}$ on the approximation ratio, while using the auxiliary graph produced by \textsc{Decomposition}$(T, 1, k)$ we get $\widehat{\theta}_p \geq k+1, p \leq t$ and $\widehat{\sigma}_p \geq \frac{k}{k+1}, p \geq t$ yielding a lower bound of $min\{\frac{e-1}{e+\frac{1}{k+1}}, \frac{k}{k+1}\}$ on the approximation ratio.
\end{lemma}
\begin{lemma}\label{lem:gen}
For any graph $G$, let $\widehat{G}$ be the auxiliary graph formed by the union of the outputs of \textsc{Decomposition}$(T, \delta, k)$, $\delta \in \{0,1\}$, on each tree $T$ of a spanning forest of $G$. Let also $\rho$ be a lower bound given by \Cref{lem:lowerbounds} (with respect to the same parameters). Then \[max\{ext(A_p), \widehat{ext}(\widehat{A}_p)\} \geq \rho\cdot ext(A_p^*).\]
\end{lemma}

By \Cref{lem:gen} it is established that \Cref{alg2} is able to guarantee a $\rho$-approximation ratio for (any instance of) Max $k$-hop Ext-Domination, where $\rho$ is a lower bound given by \Cref{lem:lowerbounds} (in fact, $\rho$ can be any other lower bound which we obtain in the same fashion as in \Cref{lem:lowerbounds}).

\subsection{A \ensuremath{(6e-5)/(6e+5)}-approximation algorithm for Max Ext-Domination}\label{subsection:improvedApprox}
We continue by showing that a more detailed analysis on \Cref{alg2} will guarantee a better lower bound on the approximation ratio for Max Ext-Domination. First, we need to better understand the inner structure of the auxiliary graph we will use. To this end, we define the following family of trees.
\begin{definition}[the $S_{n, m}$ family]\label{def:trees}
$S_{n, m}$ is a family of trees $(V, E)$ such that 
\begin{enumerate}
\item $V = \{u_0, \ldots, u_{n+m}\} \cup \{v_1, \ldots, v_m\},$
\item $E = \{\{u_0, u_1\}, \ldots, \{u_0, u_{n+m}\}\} \cup \{\{u_{n+1}, v_1\}, \ldots, \{u_{n+m}, v_m\}\}.$
\end{enumerate}
\end{definition}

We let again $G$ be a connected graph and $T$ one of its spanning trees. We also let $\widehat{G} = \bigcup_{i=1}^{t}\widehat{G}_i$ be the output of \textsc{Decomposition}($T, 1, 1$) (note that $\delta = 1$). It is easy to see that $\widehat{G}$ satisfies the following.
\begin{observation}
    $\widehat{G}_i \in S_{k_i, l_i}$, $k_i + l_i \geq 2$, for $i = 1,\ldots,t$.
\end{observation}\label{lem:D2}

Before we state the next lemma we need to clarify the assumptions we make about how \textsc{Greedy} breaks the ties. So far we have assumed that ties are broken arbitrarily. In fact, it is only in the case of the auxiliary graph that we want to prioritize the ties. First, with respect to \Cref{def:trees}, we define the \emph{center} of $\widehat{G}_i$: vertex $u_0$ is designated as the \emph{center} of $\widehat{G}_i$ unless $\widehat{G}_i \in S_{0,2}$, that is, $\widehat{G}_i$ is a path of five vertices, in which case any of the two neighbors of $u_0$ is designated as the center. The rule is the following: among equivalent vertices, any vertex that has been designated as the center of a graph $\widehat{G}_i$ is preferred over any other vertex, and a non-dominated vertex is preferred over any dominated vertex. Recall also that $(\widehat{A}_1,\ldots,\widehat{A}_p)$ is the result of \textsc{Greedy} (under the previous assumptions) when applied on the auxiliary graph $\widehat{G}$.

\begin{lemma}\label{lem:improvedRatio}
There exists an integer $\widehat{d}$ such that 
\begin{enumerate}
\item $\widehat{ext}(\widehat{A}_p) \geq \widehat{\theta}\cdot p, \text{ for }p \leq \widehat{d} \text{ where } \widehat{\theta} = \frac{6e-5}{4e-5}$,
\item $\widehat{ext}(\widehat{A}_p) \geq \widehat{\sigma}(n-p), \text{ for }p > \widehat{d} \text{ where } \widehat{\sigma} = \frac{6e-5}{6e+5}$. 
\end{enumerate}
\end{lemma}

The previous lemma allows applying \Cref{lem:gen} with $\rho = \min\{\frac{\widehat{\theta}(e-1)}{1+\widehat{\theta} e}, \widehat{\sigma}\} = \frac{6e-5}{6e+5}$ and thereby deduce that for any graph $G$,
\[
\max\{ext(A_p), ext(\widehat{A}_p)\} \geq \max\{ext(A_p), \widehat{ext}(\widehat{A}_p)\} \geq \frac{6e-5}{6e+5}ext(A_p^*),
\]
leading to the next theorem.
\begin{theorem}
\Cref{alg2} can guarantee an approximation ratio of $(6e-5)/(6e+5)$ for \textup{Max Ext-Domination}.
\end{theorem}

\section{Max Ext-Representation}
\subsection{Problem definition}
Within the context of elections, there is a winner-determination rule known as Chamberlin-Courant's voting rule~\cite{lu2011budgeted,chamberlin1983representative}. In the present work, we are interested in the approval-based variant of Chamberlin-Courant's voting rule~\cite{skowron2012fully,ijcai2022p69}. The setting of the elections, behind this rule, can in fact, be seen as an interpretation of an instance of MCP-ECC, with the rule hence coinciding with the objective of MCP-ECC. The aforementioned setting and rule are formally defined below.

\begin{definition}[The election setting]\label{def:setting}
We let $X = \{v_1, \ldots, v_n\}$ be the set of voters and $Z=\{c_1,\dots, c_m\}$ be the set of candidates. Let also $vote_i \subseteq Z$, $i = 1, \ldots, n$, denote the set of candidates which voter $v_i$ approves of, and $S_j\subseteq X$, $j = 1, \ldots, m$, denote the set of voters approving of the candidate $c_j$. We assume that each $vote_i$, $i = 1, \ldots, n$, is part of the input, while $S_j$'s can be easily obtained from $vote_i$'s.
\end{definition}

Note that, in a secrecy-preserving election, $X$ contains pseudo-identities while $Z$ contains the real identities of candidates. 

\begin{definition}[Chamberlin-Courant's voting rule]
The rule asks for electing a committee, given the votes and the number of committee members, so that the total number of represented voters, that is, voters that approve of at least one candidate on the elected committee, is maximized; the total number of represented voters is given by the following function $rep$:
\[rep(C) = \Bigg\vert\bigcup_{c_j \in C}S_{j}\Bigg\vert, C \subseteq Z.\]
\end{definition}

\paragraph*{Externally represented voters - Motivation for a new rule}
It is reasonable to argue, that the members of an elected committee are actually represented by themselves. In other words, we may say that they are \textit{internally} represented. The rest of the represented voters, on the other hand, comprise the set of \textit{externally} represented voters, that is, voters who have approved of at least one candidate on the committee, but either they are not candidates themselves or, if they are, they have not been elected. From our perspective, a natural variant of Chamberlin-Courant's voting rule would be to maximize the number of externally represented voters, as this is an actual measure of the elections' success from the point of view of the latter; this variant gives rise to the following problem.

\begin{definition}[Max Ext-Representation problem]
\textup{Max Ext-Representation} asks for electing a committee, given the votes and the number of committee members, so that the total number of externally represented voters is maximized.
\end{definition}

Before we get into further details we note that the two rules, the Chamberlin-Courant and the Max Ext-Representation, can lead to different solutions, as demonstrated by the next observation.

\begin{observation}
An optimal committee that adheres to the Chamberlin-Courant's rule is not necessarily optimal with respect to the Max Ext-Representation rule, and vice versa. To see this,
take $X = \{1, \ldots, 9\}$, $Z = \{1,2,3\}$, $S_1 = \{1, 2, 3, 4\}, S_2 = \{5, 6, 7\}, S_3 = \{1,8,9\}$ and assume that the committee must consist of two candidates (we also assume we have access into the real identities of voters). Observe that the committee $\{1,2\}$ is the only optimal solution yielded by the standard rule, while $\{2,3\}$ is the only optimal solution yielded by the variant.
\end{observation}

In order to properly address the Max Ext-Representation problem, some attention should be paid to the issue of evaluating the objective function. The difficulty stems from the fact that, in elections where voters' identities are secret, it is not possible to know whether a committee member is represented or not (since they may have voted neither for themselves nor for any other elected committee member). Therefore, we propose two settings in which the objective value of a feasible solution can be evaluated by the votes and the solution alone; we call the two settings Non-Secrecy and Rational-Candidate respectively.

\paragraph*{The Non-Secrecy Setting}
Under the Non-Secrecy Setting, voting is open ($X$ contains real identities), thus we can tell whether a participant is being internally or externally represented.
The total number of externally represented voters, with respect to a specific committee, is then given as follows:
\[ext(C) = \Bigg\vert\bigcup_{c_j \in C}S_j \setminus C\Bigg\vert, C \subseteq Z.\]

\paragraph*{The Rational-Candidate Setting}
Under the Rational-Candidate Setting, it is assumed that all candidates who vote are expected to be rational in the sense that they vote for themselves since the opposite would hurt their chances of being elected. The Rational-Candidate Setting supports secrecy, thus it may not be feasible to know who the externally represented participants are, yet, it is feasible to count them, by subtracting the number of elected candidates that are also voters; let $C_v$ denote the latter set (assume that this information is also given as part of the input, i.e., which members of $Z$ are also voters). The number of externally represented voters is then given as follows:  
\[ext(C) = rep(C) - |C_v|, C \subseteq Z.\]

\subsection{Approximation results}

\begin{observation}\label{obs:allvoters}
For any instance $\mathcal{I}$, we let $\widetilde{\mathcal{I}}$ denote a modification of $\mathcal{I}$ such that all candidates approve of themselves, including those that lack the permission of voting. The
Max Ext-Representation rule is insensitive to candidates' self-approval since it considers all committee members as internally represented, thus \[ext(C) = \widetilde{ext}(C) = \widetilde{rep}(C) - |C|.\] 
Furthermore, both Chamberlin-Courant's voting rule and the Max Ext-Representation rule yield the same solutions for $\widetilde{\mathcal{I}}$.
\end{observation}

Under the Non-Secrecy Setting, we can derive an $(e-1)/(e+1)$-approximation algorithm for Max Ext-Representation by reformulating an algorithm given in~\cite{fotakis2020object} addressing a problem of resource allocations on the vertices of an undirected graph (see \Cref{def:optext}). A suitable adaptation of the given analysis for the directed version of the latter is possible. The newly obtained results can be appropriately interpreted in conjunction with \Cref{obs:allvoters}, so as to yield the following theorem.

\begin{theorem}\label{thm:algWithMatchings}
There is an $(e-1)/(e+1)$-approximation polynomial time algorithm for \textup{Max Ext-Representation} under the Non-Secrecy Setting.
\end{theorem}

\Cref{obs:allvoters} also applies in the case of the Rational-Candidate Setting. We acknowledge that the modified instance cannot be computed, yet, its objective function can in fact be evaluated. More specifically, $\widetilde{rep}(C) = rep(C) + |C| - |C_v|$, where $C_v \subseteq C$ contains those candidates on the committee $C$ that are also voters. As already mentioned, the function $ext$ is insensitive to candidates' self-approval, however under the Rational-Candidate Setting, candidates not approving of themselves will hurt their chances of being elected, which is a motivation for all candidates to behave as expected. Admittedly, the greedy heuristic is indeed a strong candidate among the limited options offered for obtaining solid approximation results for Max Ext-Representation under the Rational-Candidate Setting. We show that under the restriction that all candidates should participate and entrust their representation to at least one other candidate other than themselves, the greedy algorithm achieves the $(e-1)/(e+1)$ approximation ratio, for both of the settings, while it remains open to determine a constant approximation ratio for the general (unrestricted) case.

\begin{theorem}\label{thm:greedyrep}
The greedy algorithm achieves an $(e-1)/(e+1)$-approximation ratio for \textup{Max Ext-Representation} under the restriction that candidates should vote for at least one candidate other than themselves. 
\end{theorem}

\begin{corollary}\label{cor:greedy}
The greedy algorithm achieves an $(e-1)/(e+1)$-approximation ratio for \textup{Max Ext-Domination}.
\end{corollary} 

\section{Applications on OPT-EXT(0,1)}
An instance of OPT-EXT(0,1) which is a special case of the more general OPT-EXT problem of graph externalities, introduced in~\cite{fotakis2020object}, consists of an undirected graph $G = (V, E)$, a collection $O$ of $m$ valued objects and a function $val: O \rightarrow \{0,1\}$ that gives the valuation $val(o)$ of every object $o \in O$. An allocation is described by a function $\pi : V(G) \rightarrow O$ that allocates all objects to individual vertices, assuming that we restrict our study to instances that satisfy $|V| = |O|$ (a $\rho$-approximate algorithm for such instances can be modified so as to yield a $\rho$-approximation guarantee for all instances~\cite{fotakis2020object}). 

\begin{definition}[OPT-EXT(0,1)]\label{def:optext}
Under an allocation $\pi$, the externality of a vertex $v \in V$, denoted as $ext_{\pi}(v)$ is either 0 or 1 and $ext_{\pi}(v) = 1$ iff $val(\pi(v)) = 0$ and there exists a vertex $u \in \mathcal{N}(v)$ such that $val(\pi(u)) = 1$; in case $ext_{\pi}(v) = 1$, we say that $v$ derives externality from $u$. The total graph externality derived under the allocation $\pi$ is given by $Ext_{\pi}(G) = \sum_{v\in V}ext_{\pi}(v)$. The \textup{OPT-EXT(0,1)} problem asks for determining an allocation $\pi$ that maximizes $Ext_{\pi}(G).$
\end{definition}

\subsection{Improved approximation algorithm for OPT-EXT(0,1)}

We let $(G, O, val)$ be an instance of OPT-EXT(0,1) and $\pi$ be an allocation. We also let $A = \{v \in V : val(\pi(v)) = 1\}$. Recall that $ext(A)$ gives the number of the vertices that are externally dominated by $A$, with respect to the graph $G$. We can observe that $Ext_{\pi}(G) \leq ext(A)$ holds. It is clear that we are only interested in allocations for which the previous relation holds with equality. We can hence deduce the following reduction and everything that comes with it, e.g., OPT-EXT(0,1) can be approximated within $(e-1)/(e+1)$ using the greedy heuristic (see \Cref{cor:greedy}) which improves on the complexity of the algorithm proposed in~\cite{fotakis2020object} that achieves the same approximation, and the next theorem that improves on the approximation ratio; the algorithm that achieves the improved approximation ratio is \Cref{alg2} which we define in more detail in \Cref{subsection:improvedApprox}, where we also show the corresponding approximation guarantees.
 \[\argmax_{\pi}Ext_{\pi}(G) \equiv \argmax_{A \subseteq V : |A| = p}ext(A),\text{ } p = \big\vert\{o \in O : val(o) = 1\}\big\vert\]

\begin{theorem}\label{thm:approxoptext01}
There exists a $(6e-5)/(6e+5)$-approximation polynomial time algorithm for \textup{OPT-EXT(0,1)}.    
\end{theorem}

What is interesting to note is that in~\cite{fotakis2020object} it is argued that OPT-EXT(0,1) does not reduce to the maximization of a monotone submodular function, as its objective as they show is neither submodular nor monotone. Nevertheless, we have shown that OPT-EXT(0,1) can actually be reduced to a problem of (non-monotone) submodular maximization.

\bibliographystyle{plain}
\bibliography{template}

\appendix

\section{Proofs from Section 2}

\subsection{Reduction graphs}
We let $G = (V, E)$ be an undirected graph and $V = \{v_1, \ldots, v_n\}$. We will show how to construct the reduction graphs $G_R^K, K \geq 2$ which will be used to establish the desired gap-preserving reductions. First, we construct the reduction graph $G_R^2$; the rest of the reduction graphs will be constructed by modifying appropriately the structure of $G_R^2$. To this end, we define a number of vertex and edge sets, as follows. We let $V^{(i)} = \{v^{(i)}_1, \ldots, v^{(i)}_n\}$ and $E^{(i)} = \big\{\{v^{(i)}_a, v^{(i)}_b\}$ $\big|$ $\{v_a, v_b\} \in E)\big\}$, for each $i = 1,\ldots,q_1$. We also let $U_i = \{u^{i}_0, \ldots, u^{i}_{q_2}\}$ and $E_i = \big\{\{u^{i}_0, u^{i}_1\}, \{u^{i}_0, u^{i}_2\}, \ldots, \{u^{i}_0, u^{i}_{q_2}\}\big\}$, for each $i = 1,\ldots,n$ (the selection of $q_1, q_2$ is explained below in a remark). Finally, we let $V' = \{v_1', \ldots, v_n'\}$ and $E' = \Big(\bigcup_{i=1}^{n}\big\{\{v_i', v^{(1)}_i\}, \{v_i', v^{(2)}_i\}, \ldots , \{v_i', v^{(q_1)}_i\}\big\}\Big) \bigcup \big\{\{v_1',u^{1}_0\}, \{v_2', u^{2}_0\}, \ldots, \{v_{n}', u^{n}_0\}\big\}$. 
\begin{remark}
The integers $q_1, q_2$ are much larger than the length of input but bounded by a polynomial of it.    
\end{remark}
\begin{remark}
Each graph $(V^{(i)}, E^{(i)})$ is isomorphic to $G$ which is to say that it is a `copy' of $G$ (hence there are $q_1$ copies of $G$) and each graph $(U_i, E_i)$ is a star of $q_2+1$ vertices with $u^{i}_0$ being its center.    
\end{remark}
The union of all the previously defined vertex sets form the vertex set of $G_R^2$ and respectively its edge set is formed by the union of all previously defined edge sets. To construct $G_R^K, K > 2$ we remove every edge $\{v_i', u^{i}_0\}$ and replace it with a path of $K-1$ edges which includes both $v_i', u^{i}_0$. Note that $dist(v_i', u^{i}_0) = K-1$, for $i = 1,\ldots,n$.

\subsection{Gap-preserving reductions}
Let $OPT_{(p,K)}(G_R^K)$ denote the maximum number of vertices that a set of $p$ vertices of $G_R^K$ can dominate (recall that a vertex can dominate all vertices that are in a distance of at most $k$ hops, including themselves).

\begin{lemma} It holds,
\begin{enumerate}
\item If $OPT_{(p,K-1)}(G) = $ \textbf{\textup{max}}, then $OPT_{(p,K)}(G_R^K) \geq p(q_2+K) + q_1\cdot\textbf{\textup{max}}$
\item If $OPT_{(p,K-1)}(G) \leq (1-\frac{1}{e})$\textbf{\textup{max}}, then 

\[OPT_{(p,K)}(G_R^K) \leq  (1-\frac{1}{e} + \frac{p(q_2+K) + e(K-2)(n-p)}{e(p(q_2+K) + q_1\cdot\textbf{\textup{max}})})(p(q_2+K) + q_1\cdot\textbf{\textup{max}})
\]
\end{enumerate}
\end{lemma}
\begin{proof}
~\begin{enumerate}
\item Assuming that $OPT_{(p,K-1)}(G)$ is derived from the optimal set of vertices $\{v_{i_1}, \ldots, v_{i_p}\}$ then the number of vertices of $G_R^K$ that the set $\{v_{i_1}', \ldots, v_{i_p}'\}$ dominates, is $p(q_2+K) + q_1\cdot\textbf{max} + \alpha(K-2)(n-p)$, $0\leq\alpha \leq 1$. Therefore $OPT_{(p,K)}(G_R^K) \geq p(q_2+K) + q_1\cdot\textbf{max}$.
\item Let $u$ be a vertex of any path from $v_i'$ to the leaves of the $i$-th star. It is easy to see that $\mathcal{N}_K[u] \subseteq \mathcal{N}_K[v_i']$, $i \in [1,n]$. Furthermore, $\big|\mathcal{N}_K(v_i') - \mathcal{N}_K(v^{(j)}_i)\big| \geq q_2$ while $\big|\mathcal{N}_K(v^{(j)}_i) - \mathcal{N}_K(v_i')\big| < 2n, (i, j) \in [1,n] \times [1,q_1]$. By considering $q_2 \geq 2n$, we know that there must be an optimal set of $p$ vertices of $G_R^K$ which is a subset of $V'$ thus the derived optimal solution is $OPT_{(p,K)}(G_R^K) = p(q_2+K) + \alpha(K-2)(n-p) + q_1\cdot M$, where $0 \leq \alpha \leq 1$ and $M$ is the number of vertices of any graph $G^{(i)}$ that the optimal set dominates. If $OPT_{(p,K-1)}(G) \leq (1-\frac{1}{e})$\textbf{max} then $OPT_{(p,K)}(G_R^K) \leq p(q_2+K) + (K-2)(n-p) + q_1(1-\frac{1}{e})\textbf{max} \Rightarrow
OPT_{(p,K)}(G_R^K) \leq (1-\frac{1}{e} + \frac{p(q_2+K) + e(K-2)(n-p)}{e(p(q_2+K) + q_1\cdot\textbf{max})})(p(q_2+K) + q_1\cdot\textbf{max})$.
\end{enumerate}   
\end{proof}

\subsection{Proof of Theorem 1}
Let $\mathcal{R}_K$ be a polynomial time algorithm which on input $G$ outputs the graph $G_R^K$, for $K \geq 2$. Let $q_1^{(K)}, q_2^{(K)}$ be the corresponding integers. We have $|V(G_R^K)| = (q_1^{(K)} + q_2^{(K)} + K)n$, where $n = |V(G)|$. Now consider the following graph: $\mathcal{R}_{j}(\mathcal{R}_{j-1}(\ldots\mathcal{R}_2(G)\ldots))$, which we denote as $\mathcal{G}_j, 2 \leq j \leq K$. It easily follows that
\[|V(\mathcal{G}_j)| = (q_1^{(j)} + q_2^{(j)} + j)|V(\mathcal{G}_{j-1})|\]
Furthermore, we know that $q_1^{(j)}$ is a polynomial of $|V(\mathcal{G}_{j-1})|$ , while $q_2^{(j)}$ is a linear polynomial of $|V(\mathcal{G}_{j-1})|$.
Hence,
\[|V(\mathcal{G}_j)| = \Theta(|V(\mathcal{G}_{j-1})|^{c_{j-1}}), \text{ for some constant $c_{j-1}$ }\]
Ultimately, there exists a constant $c = \prod_{i=1}^{K-1}c_i$ (which does not depend on $n$) such that
\[|V(\mathcal{G}_K)| = \Theta(n^c)\]
Using the previous lemma, a gap-preserving reduction can be established in polynomial time from \textup{Max Domination} to \textup{Max $k$-hop Domination} for any fixed $k \geq 2$. Therefore, no $(1-1/e + \varepsilon)$-approximation polynomial time algorithm exists for \textup{Max $k$-hop Domination} for any $\varepsilon > 0$, unless \textup{\textbf{P} = \textbf{NP}}.

\section{Proofs from Section 3}

\subsection{Proof of Proposition 1}
Note that we can write $dom(A_p) - p = \sum_{i=1}^{p}(dom(A_i)-dom(A_{i-1})-1)$. We let $w_i = dom(A_i)-dom(A_{i-1})-1$ and $v_{\alpha_i} = \argmax_{v \in V\backslash A_{i-1}}dom(A_{i-1} \cup \{v\}), i=1,\ldots,p$. We have, \[dom(A_{i-1} \cup \{v_{\alpha_{i}}\}) - dom(A_{i-1}) \geq dom(A_{i-1} \cup \{v_{\alpha_{i+1}}\}) - dom(A_{i-1}),\] 
while the submodularity of $dom$ allows writing \[dom(A_{i-1} \cup \{v_{\alpha_{i+1}}\}) - dom(A_{i-1}) \geq dom(A_i \cup \{v_{\alpha_{i+1}}\}) - dom(A_i).\] Combining the previous two inequalities we get $w_i \geq w_{i+1}$. Thus,
\begin{multline*}
p\cdot w_1 + \ldots + p\cdot w_p + p\cdot w_{p+1} \leq p\cdot w_1 + \ldots + p\cdot w_p + (w_1 + \ldots +  w_p) \\ \Rightarrow p\sum_{j=1}^{p+1}w_j \leq (p+1)\sum_{j=1}^{p}w_j \Rightarrow \frac{ext(A_{p+1})}{p+1} \leq \frac{ext(A_p)}{p}.
\end{multline*}

Furthermore, there exists an integer $p'$ such that $dom(A_1) < \ldots < dom(A_{p'})$ and $dom(A_{p}) = n$, for $p > p'$. Consequently, $ext(A_1) \leq \ldots \leq ext(A_{p'})$ and $ext(A_p) = n-p$, for $p > p'$. Therefore, $\sigma_1 \leq \ldots \leq \sigma_{p'} \leq 1$ and $\sigma_p = 1$, for $p > p'$.

\subsection{Proof of Lemma 1}

For $\delta = 0$ we directly deduce that,
\[\widehat{ext}(\widehat{A}_p) \geq k\cdot p, p \leq t \Rightarrow \widehat{\theta}_p \geq k, p \leq t\]
and
\[\widehat{ext}(\widehat{A}_p) = n-p, p \geq t \Rightarrow \widehat{\sigma}_p = n-p, p\geq t,\]
yielding a lower bound of $\frac{e-1}{e+\frac{1}{k}}$ on the approximation ratio.

While for $\delta = 1$, we observe first that the height of $u_i$ is at most $k+1$ because $\widehat{G}_i$ is minimal. If the height of $u_i$ is at most $k$ then $u_i$ can dominate all its descendants (since they are within $k$ hops from $u_i$). Otherwise, if the height of $u_i$ is $k+1$ then every subtree of $\widehat{G}_i$ with height equal to $k$ is, in fact, a path of $k+1$ vertices. Regarding the number of vertices that can be externally dominated in the latter case where the height of $\widehat{G}_i$ is $k+1$, we distinguish the following two cases:
\begin{enumerate}
\item Assuming $\widehat{G}_i$ is a path of $k+2$ vertices, any internal vertex (except the root) can dominate the rest of the vertices.
\item Assuming the subtrees formed by the children of $u_i$ form the graph $X_i \cup Y_i$, where $X_i$ is a union of $\lambda_i$ paths of $k+1$ vertices each and $Y_i$ is the union of the rest of the subtrees which are of a height of at most $k-1$, vertex $u_i$ externally dominates $\lambda_i\cdot k + |V(Y_i)| \geq k + 1$ vertices. Observe that $\frac{\lambda_i\cdot k + |V(Y_i)|}{|V(\widehat{G}_i)|-1} = \frac{\lambda_i\cdot k + |V(Y_i)|}{\lambda_i\cdot(k+1) + |V(Y_i)|} \geq \frac{k}{k+1}$.
\end{enumerate}
It follows that
\[\widehat{ext}(\widehat{A}_p) \geq (k+1)p, p \leq t \Rightarrow \widehat{\theta}_p \geq k+1, p \leq t\]
and
\[\widehat{ext}(\widehat{A}_t) \geq \frac{k}{k+1}\sum_{j=1}^{t}(|V(\widehat{G}_j)|-1) = \frac{k}{k+1}(|V(\widehat{G})|-t)\]   \[\Rightarrow \widehat{\sigma}_t \geq \frac{k}{k+1} \Rightarrow \widehat{\sigma}_p \geq \frac{k}{k+1}, p \geq t,\]
guaranteeing a lower bound of $\min\{\frac{e-1}{e+\frac{1}{k+1}}, \frac{k}{k+1}\}$ on the approximation ratio.

\subsection{Proof of Lemma 2}
We let $G$ be any graph, not necessarily connected. We also let $G_1$ be the union of all its connected components, each having at least $k+1+\delta 
$ $(\delta \in \{0,1\}) $ vertices and $G_2$ be the union of all the remaining connected components. We let $dom_G, dom_{G_1}, dom_{G_2}$ denote the standard objective functions with respect to the graphs $G, G_1, G_2$. We consider three instances each associated with one of the graphs $G, G_1, G_2$, where the number of dominators in each instance is $p, p_1, p_2$ respectively, with $p_1 + p_2 = p$. We let $(A^{G}_1, \ldots, A^{G}_p), (A^{G_1}_1, \ldots, A^{G_1}_{p_1}), (A^{G_2}_1, \ldots, A^{G_2}_{p_2})$ be the results of greedy algorithm for each instance and $A^{G *}_p, A^{G_1 *}_{p_1}, A^{G_2 *}_{p_2}$ be the respective optimal solutions. We assume that same ties in different instances break in the same way. By the definition of the greedy algorithm, we immediately get \[dom_G(A^{G}_p) \geq dom_{G_1}(A^{G_1}_{p_1}) + dom_{G_2}(A^{G_2}_{p_2})\]
which implies that \[ext_G(A^{G}_p) \geq ext_{G_1}(A^{G_1}_{p_1}) + ext_{G_2}(A^{G_2}_{p_2})\]
Observe that we can decompose each tree of the spanning forest of the graph $G_1$ in the same way we decompose the spanning tree of any connected graph (of at least $k+1+\delta$ vertices). Let $\widehat{G}$ be the union of the decomposed spanning forest $\widehat{G}_1$ (of $G_1$) and the graph $G_2$. Similarly, we define the functions $\widehat{dom}_{\widehat{G}}, \widehat{dom}_{\widehat{G}_1}$ and the sequences $(A^{\widehat{G}}_1, \ldots, A^{\widehat{G}}_p), (A^{\widehat{G}_1}_1, \ldots, A^{\widehat{G}_1}_p)$. We also have,
\[\widehat{ext}_{\widehat{G}}(A^{\widehat{G}}_p) \geq \widehat{ext}_{\widehat{G}_1}(A^{\widehat{G}_1}_{p_1}) + ext_{G_2}(A^{G_2}_{p_2})\]
Hence, 
\begin{align*}\max\{ext_G(A^{G}_p), \widehat{ext}_{\widehat{G}}(A^{\widehat{G}}_p) \} \geq \max\{ext_{G_1}(A^{G_1}_{p_1}), \widehat{ext}_{\widehat{G}_1}(A^{\widehat{G}_1}_{p_1})\} + ext_{G_2}(A^{G_2}_{p_2})
\end{align*}
Let $\rho$, $0 < \rho \leq 1$ be such that for any graph $G_1$ (i.e. a graph, every connected component of which has at least $k+1+\delta$ vertices), and the corresponding decomposed graph $\widehat{G}_1$, \[\max\{ext_{G_1}(A^{G_1}_{p_1}), \widehat{ext}_{\widehat{G}_1}(A^{\widehat{G}_1}_{p_1})\} \geq \rho \cdot ext_{G_1}(A^{G_1 *}_{p_1}),\]
holds.

Observe that \Cref{lem:lowerbounds} also applies to graph $G_1$, thus we can obtain such number $\rho$ that satisfies the above relation, using \cref{lem:lowerbounds}.

Furthermore, we can easily show that $ext_{G_2}(A^{G_2}_{p_2}) = ext_{G_2}(A^{G_2 *}_{p_2})$; since every connected component of $G_2$ has at most $k+\delta$ vertices it follows that any vertex can dominate all of the rest of the vertices (within the connected component). 
Thus,
\[\max\{ext_G(A^{G}_p), \widehat{ext}_{\widehat{G}}(A^{\widehat{G}}_p)\} \geq \rho\cdot (ext_{G_1}(A^{G_1 *}_{p_1}) + ext_{G_2}(A^{G_2 *}_{p_2}))\]
If we take $p_1 = |A^{G *}_p \cap V(G_1)|$ and $p_2 = p - p_1$, we get
\begin{align*}\max\{ext_G(A^{G}_p), ext_G(A^{\widehat{G}}_p)\} \geq \max\{ext_G(A^{G}_p), \widehat{ext}_{\widehat{G}}(A^{\widehat{G}}_p)\} \geq \rho\cdot ext_{G}(A^{G *}_{p}). 
\end{align*}

\subsection{Proof of Lemma 3}
For the sake of simplicity, we abuse notation and write $ext(p, G)$ to denote the evaluation of the function $ext$ on the output produced by \textsc{Greedy}($G, p, 1$). 

Let $G$ be a connected graph and $\widehat{G}$ be the auxiliary graph produced by \textsc{Decomposition}($T, 1, 1$) with $T$ being any spanning tree of $G$. Let $H$ be the subgraph of $\widehat{G}$ formed by the union of all ($q$ in number) connected components $\widehat{G}_i$ such that $\widehat{G}_i\in S_{0,2}$ and $H'$ the subgraph of $\widehat{G}$ consisting of the union of all ($q'$ in number) connected components of $\widehat{G}_i$ such that $\widehat{G}_i \in S_{0,3}$. 

We can wlog denote $\widehat{G} = \Big\{\widehat{G}_1, \ldots, \widehat{G}_{t-q-q'}\Big\} \bigcup H \bigcup H'$. We will derive lower bounds for each of the three subgraphs comprising the graph $G$ appearing in the union above and the lower bound with respect to $G$ will follow. 

To begin with, we define the following numbers $x_i$ each associated with the respective component $\widehat{G}_i$, for $1 \leq i \leq t - q - q'$. 
\[x_i =
\left\{
	\begin{array}{lll}
		1  & \mbox{if } l_i \in \{0,1\} \\
		\floor*{\frac{l_i+1}{2}} & \mbox{if } k_i \geq 1,  l_i \geq 2 \\
		\floor*{\frac{l_i}{2}} & \mbox{if } k_i = 0, l_i \geq 4 \\
	\end{array}
\right\}\]
Let $n_i = \Big|V(\widehat{G}_i)\Big|$, $i \leq t$. Now, observe that the following hold, regarding $ext(x_i, \widehat{G}_i):$
\begin{enumerate}
\item If $l_i = 0$ then
$ext(1, \widehat{G}_i) = k_i \geq 2$ and $ext(1,\widehat{G}_i) = n_i-1$
\item If $l_i = 1$ then
$ext(1, \widehat{G}_i) = k_i + 1 \geq 2$ and $ext(1, \widehat{G}_i) \geq  \frac{2}{3}(n_i - 1)$
\item If $k_i \geq 1, l_i \geq 2$ then
$ext(\floor*{\frac{l_i+1}{2}}, \widehat{G}_i) = k_i + l_i \geq 2\floor*{\frac{l_i+1}{2}}$. Furthermore, $\frac{k_i + l_i}{k_i + 2l_i+1-\floor*{\frac{l_i+1}{2}}} \geq \frac{k_i+l_i}{1.5(k_i+l_i)+0.5} \geq \frac{3}{5} \Rightarrow ext(\floor*{\frac{l_i+1}{2}}, \widehat{G}_i) \geq \frac{3}{5}(n_i-\floor*{\frac{l_i+1}{2}})$.
\item If $k_i = 0, l_i \geq 4$ then
$ext(\floor*{\frac{l_i}{2}}, \widehat{G}_i) = l_i$. Furthermore, for even $l_i$ we have $\frac{l_i}{2l_i + 1 - \floor*{\frac{l_i}{2}}} = \frac{l_i}{1.5l_i + 1} \geq \frac{4}{7} \Rightarrow ext(\floor*{\frac{l_i}{2}}, \widehat{G}_i) \geq \frac{4}{7}(n_i-\floor*{\frac{l_i}{2}})$. While, for odd $l_i$ we have, $\frac{l_i}{2l_i+1-\floor*{\frac{l_i}{2}}} = \frac{l_i}{1.5(l_i+1)} \geq \frac{5}{9} \Rightarrow ext(\floor*{\frac{l_i}{2}}, \widehat{G}_i) \geq \frac{5}{9}(n_i-\floor*{\frac{l_i}{2}})$.
\end{enumerate}

Let $x = \sum_{i=1}^{t-q-q'}x_i$. For $p \leq t-q-q'$, it is clear that $ext(p, \bigcup_{i=1}^{t-q-q'}\widehat{G}_i) \geq 2p$ and for $t-q-q' < p \leq x,$ 
\begin{multline*}ext(p, \bigcup_{i=1}^{t-q-q'}\widehat{G}_i) = \sum_{i=1}^{t-q-q'}(k_i+l_i) = \sum_{i=1}^{t-q-q'}ext(x_i, \widehat{G}_i) \geq \sum_{i=1}^{t-q-q'}2x_i = 2x \\ \Rightarrow ext(p, \bigcup_{i=1}^{t-q-q'}\widehat{G}_i) \geq 2p
\end{multline*}
Furthermore, 
\begin{multline*}
ext(x, \bigcup_{i=1}^{t-q-q'}\widehat{G}_i) = \sum_{i=1}^{t-q-q'}(k_i+l_i) = \sum_{i=1}^{t-q-q'}ext(x_i, \widehat{G}_i) \geq \sum_{i=1}^{t-q-q'}\frac{5}{9}(n_i-x_i) \\ \Rightarrow ext(x, \bigcup_{i=1}^{t-q-q'}\widehat{G}_i) \geq \frac{5}{9}((n-|V(H)|-|V(H')|)-x).
\end{multline*}
We thus have \[ext(p, \bigcup_{i=1}^{t-q-q'}\widehat{G}_i) \geq 2p, p \leq x \text{ and } ext(p, \bigcup_{i=1}^{t-q-q'}\widehat{G}_i) \geq \frac{5}{9}((n-|V(H)|-|V(H')|)-p), p \geq x.\]

Next, by taking $\alpha = \frac{\floor*{c\cdot q}}{q}$, $0 \leq c < 1$, we can show that \[ \frac{ext((1+\alpha)q,H)}{(1+\alpha)q} = \frac{(2+\alpha)q}{(1+\alpha)q} \geq \frac{2+c}{1+c}\] 
and 
\[\frac{ext((1+\alpha)q+1,H)}{|V(H)|-((1+\alpha)q+1)} = \frac{(2+\alpha)q+1}{5q-(1+\alpha)q-1} = \frac{2q+(\alpha q+1)}{4q-(\alpha q+1)} > \frac{2q+c\cdot q}{4q-c\cdot q} = \frac{2+c}{4-c}.\] Observe that
\[\frac{\frac{2+c}{1+c}(e-1)}{1+\frac{2+c}{1+c}e} = \frac{2+c}{4-c} \Rightarrow c = 1-\frac{5}{2e}.\]
Thus, for $y = q + \floor*{c\cdot q}$,  \[ext(p, H) \geq \frac{6e-5}{4e-5}p, p \leq y \text{ and } ext(p, H) \geq \frac{6e-5}{6e+5}(|V(H)|-p), p\geq y+1.\]

In the same fashion, for $\beta = \frac{\floor*{\frac{q'}{2}}}{q'}$, \[ext((1+\beta)q', H') = 3q' \geq 2(1+\beta)q'\] 
and 
\[\frac{ext((1+\beta)q'+1,H')}{|V(H')|-((1+\beta)q'+1)} = \frac{3q'}{7q' - (1+\beta)q'-1} = \frac{3q'}{6q'-(\beta q' + 1)} > \frac{3q'}{6q'-\frac{q'}{2}} = \frac{6}{11}.\]

Thus for $y' = q' + \floor*{\frac{q'}{2}}$, \[ext(p, H') \geq 2p, p \leq y' \text{ and } ext(p, H') \geq \frac{6}{11}(|V(H')|-p), p \geq y'+1.\] 

Further observe that if $q' \notin \{1,3\}$ then
\begin{equation*}
\frac{ext(y', H')}{|V(H')|-y'} =  \frac{3q'}{7q'-(1+\beta)q'} = \frac{3q'}{(6-\beta)q'} \geq \frac{15}{28},
\end{equation*}
deducing that 
\[ext(p, H') \geq 2p, p \leq y' \text{ and } ext(p, H') \geq \frac{15}{28}(|V(H')|-p), p \geq y',\]
while if $q' \in \{1, 3\}$, by taking random $\widehat{G}_j \in H$ we have
\[
\frac{ext(y'+2, H' \cup \widehat{G}_j)}{|V(H' \cup \widehat{G}_j)|-(y'+2)} = \frac{3q'+3}{(6-\beta)q'+3} \geq \frac{3}{5}, \]
deducing that \[ext(p, H' \cup \widehat{G}_j) \geq 2p, p \leq y' + 2 \text{ and } ext(p, H' \cup \widehat{G}_j) \geq \frac{3}{5}(|V(H' \cup \widehat{G}_j)|-p), p \geq y'+2.\]

Finally, if we determine $d = x + y + y'$ for $q' \notin \{1, 3\}$ and $d = x + y + y' + 2$ otherwise, we deduce the desired, that is,
\[ext(p, G) \geq \frac{6e-5}{4e-5}p, p \leq d \text{ and } ext(p, G) \geq \frac{6e-5}{6e+5}(n-p), p > d.\]

\section{Proofs from Section 4}
\subsection{Proof of Theorem 3}
We will only present those steps that need to be adapted for our case. To this end, we let $A_p, A_p^*$ be the greedy and the optimal solution respectively for any modified instance as described by \Cref{obs:allvoters}. With respect to $A_p^*$, we will create a sequence $C_1, ..., C_p$ of $p$ sets, each containing exactly one distinct elected candidate, i.e., a candidate which appears in $A_p^*$, and a number of represented voters, iteratively in $p$ steps as follows: At each step, we choose the elected candidate that can externally represent the maximum number of voters which are not (externally) represented by any previously chosen candidate. The $i$-th set consists of the aforementioned candidate and the corresponding externally represented voters. We let $t^*$ be the maximum index so that $|C_{t^*}| \geq 2$ (note that $|C_{i}| \geq |C_{i+1}|$).
Observe that,
\[ext(A_p^*) = \Bigg|\bigcup_{i=1}^{t^*}C_i\Bigg| - t^*.\]
Furthermore, we can wlog assume that,
\[ext(A_p) \geq ext(A_{t^*}).\]
Note that the opposite would mark $A_p$ as an optimal solution ($A_p$ would then 
externally represent every voter, including non-elected candidates). 
We also have,
\[\widetilde{rep}(A_{t^*}) \geq (1-1/e)\Bigg|\bigcup_{i=1}^{t^*}C_i\Bigg|.\]
The combination of the previous two inequalities gives,
\[ext(A_p) \geq (1-1/e)\Bigg|\bigcup_{i=1}^{t^*}C_i\Bigg| - t^*.\]
The rest of the steps require computing a maximum matching in order to derive an auxiliary solution. The problem of object allocations considers undirected graphs. In our case, we can represent the elections with the help of a directed graph. Then a definition for a maximum matching that would fit our purposes is given as follows: We transform every directed edge into an undirected one, remove any duplicate edges that may appear, and then compute a maximum matching for the yielded (undirected) graph. By the construction of the sequence $C_1, ..., C_p$ we do know that a matching of $t^*$ arcs is guaranteed. A maximum matching, as defined above, can be thus used to derive the auxiliary solution where at least $t^*$ voters will be externally represented. This completes the adaptation effort and the rest of the proof produces the desired: among the two committees, i.e., the committee yielded by the greedy algorithm and the committee yielded with respect to the maximum matching, the algorithm that outputs the committee that achieves the maximum number of externally represented voters, achieves an approximation ratio of $(e-1)/(e+1)$.

\subsection{Proof of Theorem 4}
We let $A_p, A_p^* \subseteq Z$ be the solution of the greedy algorithm and the optimal solution respectively for a given instance (see \Cref{obs:allvoters}). Let also $p'$ be the largest index such that $ext(A_1) < \ldots < ext(A_{p'})$, which implies that $ext(A_p) \geq p$, $p \leq p'$. We know that $\widetilde{rep}(A_p) \geq (1-1/e)\widetilde{rep}(A_p^*)$ which we rewrite as $ext(A_p) \geq (1-1/e)ext(A_p^*) - p/e$ and since $ext(A_p) \geq p$, we can deduce that \[ext(A_p) \geq \frac{e-1}{e+1}ext(A_p^*), p \leq p'\]

The candidates who are not elected, considering $A_{p'}$ as the elected committee, can be divided into two categories described by the following two observations.

\begin{observation}\label{rep:obs1}
Let $c_j$ be a candidate who does not approve of anyone on the committee $A_{p'}$. Then every voter in $S_j \backslash \{c_j\}$ approves of at least one candidate on the committee.    
\end{observation}
\begin{observation}\label{rep:obs2}
Let $c_j$ be a candidate who is not on the committee $A_{p'}$ but approves of at least one candidate (on the committee). Then, at most one voter in $S_j$ does not approve of anyone on the committee.    
\end{observation}

Now, assume that $A_p^* = \{i_1, \ldots, i_\lambda\} \cup \{j_1, \ldots, j_\mu\}$, where $\lambda + \mu = p \geq p'$, $\{i_1, \ldots, i_\lambda\} \subseteq A_{p'}$ and $\{j_1, \ldots, j_\mu\} \cap A_{p'} = \emptyset$. Let $Q, Q_1, Q_2$ be the voters that approve of at least one candidate in $A_{p'}, \{i_1, \ldots, i_\lambda\}, \{j_1, \ldots, j_\mu\}$ respectively.
We have, $ext(A_p^*) = (|Q_1| - \lambda) + (|Q_2\cap Q - Q_1| + |Q_2 - Q| - \mu) = (|Q_1| - \lambda) + (|Q_2\cap Q - Q_1| - \mu) + |Q_2 - Q| \leq (|Q_1| - \lambda) + (|Q - Q_1| - \mu) + |Q_2 - Q| \leq ext(A_{p'}) + |Q_2 - Q|$.
\begin{observation}\label{rep:obs3}
If every candidate approves of at least one candidate other than themselves, then for every candidate $c_j$ described by \Cref{rep:obs1} there exists a candidate $c_{j'}$ described by \Cref{rep:obs2} such that $c_j \in S_{j'}$. 
\end{observation}

Should a committee $C_p$ of $p$ candidates could be computed so that at least $|Q_2 - Q|$ voters are externally represented then the maximum between the total number of voters that are externally represented by $C_p$ and $A_p$ respectively would guarantee an $1/2$-approximation ratio. First, we examine whether $A_p$ would fit. We will show that indeed $ext(A_p) \geq |Q_2 - Q|$.  Let $\{j_{\alpha_1}, \ldots, j_{\alpha_\nu}\} \subseteq \{j_1, \ldots, j_{\mu}\}$, $\nu \leq \mu$ such that each candidate $j_{\alpha_i}$ is described by \Cref{rep:obs2}. Using \Cref{rep:obs3} we can deduce that $Q_2 - Q$ consists of exactly those voters that approve of at least one candidate in $\{j_{\alpha_1}, \ldots, j_{\alpha_\nu}\}$ and also that $|Q_2 - Q| \leq \nu$. Furthermore, we have $ext(A_p) \geq \nu$ hence, $ext(A_p) \geq \frac{1}{2}ext(A_p^*)$, $p \geq p'$. We acknowledge that we have implicitly assumed that $ext(A_p) = ext(A_{p'})$ but it is trivial to observe that if $ext(A_p) < ext(A_{p'})$ then all voters are represented by at least one candidate on the committee $A_p$, marking $A_p$ as an optimal committee, which completes the proof.

\subsection{Proof of Corollary 4.1}
Max Ext-Domination as already pointed out is a special case of the Max Ext-Representation problem (under the Non-Secrecy Setting), and the assumption we have made for the latter is satisfied for Max Ext-Domination if there are no isolated vertices (if there are, the algorithm can be trivially adapted to yield the same ratio).







\end{document}